\documentstyle[buckow]{article}
\input epsf
\newcommand{\NPB}[3]{\emph{ Nucl.~Phys.} \textbf{B#1} (#2) #3}   
\newcommand{\PLB}[3]{\emph{ Phys.~Lett.} \textbf{B#1} (#2) #3}   
\newcommand{\PRD}[3]{\emph{ Phys.~Rev.} \textbf{D#1} (#2) #3}

\newcommand{\JHEP}[3]{\emph{JHEP} \textbf{#1} (#2) #3}
\newcommand{\bea}{\begin{eqnarray}}
\newcommand{\eea}{\end{eqnarray}}
\newcommand{\be}{\begin{equation}}
\newcommand{\ee}{\end{equation}}
\begin {document}
\def\titleline{D-brane Standard Model\footnote{Presented by E. Kiritsis}
}
\def\authors{I. Antoniadis\1ad\2ad, E. Kiritsis\3ad and T. Tomaras\3ad
}
\def\addresses{\1ad Theory Division, CERN, CH-1211, Geneva 23, Switzerland\\
\2ad Centre de Physique Th{\'e}orique, Ecole Polytechnique, 91128 Palaiseau, France\\
\3ad Department of Physics, University of Crete, and FO.R.T.H., 
P.O. Box 2208, 710 03 Heraklion, Greece
}
\def\abstracttext{
The minimal embedding of the Standard Model in type I string theory is described.
The $SU(3)$ color and $SU(2)$ weak
interactions arise from two different collections of branes.
The correct prediction of the weak angle is obtained for a string scale of 6-8 TeV.
Two Higgs doublets are necessary and proton stability is guaranteed.
It predicts two massive vector bosons with masses at the TeV scale, as well as a new superweak
interaction.
}
\large
\makefront
\section{Introduction}
String theory is the only known framework for quantizing gravity. If its
fundamental scale is of the order of the Planck mass, stability of the 
hierarchy of the weak scale requires low energy supersymmetry. This framework fits nicely
with the apparent unification of the gauge couplings in the minimal
supersymmetric standard model. 
However, breaking supersymmetry at low energies is a hard problem, which in
string perturbation theory implies a large extra dimension, \cite{bac,k,ant}.
Recently, an alternative approach has been put
forward \cite{add,aadd} in which stabilization of the hierarchy is achieved without
supersymmetry, by lowering the string scale down to a few TeV
\cite{ant,Ant,aadd,twoc,st}. A
natural realization of this possibility is offered by weakly coupled type I string
theory, where gauge interactions are described by open strings whose ends are
confined on D-branes, while gravity is mediated by closed strings in the bulk
\cite{aadd}. The observed hierarchy between the Planck and the weak scales is then
accounted for by two or more large dimensions, transverse to our brane-world, with
corresponding size varied from a millimeter to a fermi.

One of the main questions with such a low string scale is to understand the
observed values of the low energy gauge couplings. One possibility is to have
the three gauge group factors of the Standard Model arising from different
collections of coinciding branes. This is unattractive since the three gauge
couplings correspond in this case to different arbitrary parameters of the
model. A second possibility is to maintain unification by imposing all the
Standard Model gauge bosons to arise from the same collection of D-branes. The
large difference in the actual values of gauge couplings could then be explained
either by introducing power-law running from a few TeV to the weak scale
\cite{ddg}, or by an effective logarithmic evolution in the transverse space in
the special case of two large dimensions \cite{abd}. However, no satisfactory model
built along these lines has so far been presented.

A third possibility exists \cite{akt} which is alternative to 
unification
but nevertheless maintains the prediction of the weak angle at low energies.
Specifically, we consider  the strong and electroweak interactions to 
arise from two different
collections of coinciding branes, leading to two different gauge couplings,
\cite{twoc}.
Assuming that the low energy spectrum of the (non-supersymmetric) Standard Model
can be derived by a type I/I$^\prime$ string vacuum, the normalization of the
hypercharge is determined in terms of the two gauge couplings and leads
naturally to the right value of $\sin^2\theta_W$ for a string scale of the order
of a few TeV. The electroweak gauge symmetry is broken by the vacuum expectation
values of two Higgs doublets, which are both necessary in the present context to
give masses to all quarks and leptons.

Another issue of this class of models with TeV string scale is to understand
proton stability. In the model presented here, this is achieved by the
conservation of the baryon number which turns out to be a perturbatively
exact global symmetry, remnant of an anomalous $U(1)$ gauge symmetry broken by
the Green-Schwarz mechanism. Specifically, the anomaly is canceled by shifting a
corresponding axion field that gives mass to the $U(1)$ gauge boson.

As it turns out, some of the standard model fermions are fluctuations of strings ending in a
different brane. This implies the presence of a  new short range 
interaction due to the gauge bosons
and/or scalars representing the fluctuations of the extra brane.
Moreover, the two extra U(1) gauge groups are anomalous and the associated gauge bosons become massive 
with  masses of the order of the string scale.
Their couplings to the standard model fields up to dimension five are fixed by charges and anomalies.

\section{Hypercharge embeddings and the weak angle}

The gauge group closest to the $SU(3)\times SU(2)\times U(1)$ of the Standard
Model one can hope to derive from type I/I$^\prime$ string theory in the above
context is $U(3)\times U(2)\times U(1)$. The first factor arises from three
coincident D-branes (``color" branes). An open string with one end on them is a
triplet under $SU(3)$ and carries the same $U(1)$ charge for all three components.
Thus, the $U(1)$ factor of $U(3)$ has to be identified with {\it gauged} baryon
number. Similarly, $U(2)$ arises from two coincident ``weak" D-branes and the
corresponding abelian factor is identified with {\it gauged} weak-doublet
number. A priori, one might expect that $U(3)\times U(2)$ would be the minimal
choice. However it turns out that one cannot give masses to both up and down quarks in that case.
Therefore,
at least one additional $U(1)$ factor corresponding to an extra D-brane
(``$U(1)$" brane) is necessary in order to accommodate the Standard 
Model. In principle this $U(1)$ brane can be chosen to be independent of the other
two collections with its own gauge coupling. To improve the predictability of the
model, here we choose to put it on top of either the color or the weak D-branes. In
either case, the model has two independent gauge couplings
$g_3$ and $g_2$ corresponding, respectively, to the gauge groups $U(3)$ and
$U(2)$. The $U(1)$ gauge coupling $g_1$ is equal to either $g_3$ or $g_2$.

Let us denote by $Q_3$, $Q_2$ and $Q_1$ the three $U(1)$ charges of $U(3)\times
U(2)\times U(1)$, in a self explanatory notation. Under $SU(3)\times SU(2)\times
U(1)_3\times U(1)_2\times U(1)_1$, the members of a family of quarks and
leptons have the following quantum numbers:
\bea
&Q &({\bf 3},{\bf 2};1,w,0)_{1/6}\nonumber\\
&u^c &({\bf\bar 3},{\bf 1};-1,0,x)_{-2/3}\nonumber\\
&d^c &({\bf\bar 3},{\bf 1};-1,0,y)_{1/3}\label{charges}\\
&L   &({\bf 1},{\bf 2};0,1,z)_{-1/2}\nonumber\\
&l^c &({\bf 1},{\bf 1};0,0,1)_1\nonumber
\eea
Here, we normalize all $U(N)$ generators according to
${\rm Tr}T^aT^b=\delta^{ab}/2$, and measure the corresponding $U(1)_N$ charges
with respect to the coupling $g_N/\sqrt{2N}$, with $g_N$ the $SU(N)$ coupling
constant. Thus, the fundamental representation of $SU(N)$ has $U(1)_N$ charge
unity. The values of the $U(1)$ charges $x,y,z,w$ will be fixed below so that
they lead to the right hypercharges, shown for completeness as subscripts.

The quark doublet $Q$ corresponds necessarily to a massless excitation of an
open string with its two ends on the two different collections of branes. The
$Q_2$ charge $w$ can be either $+1$ or $-1$ depending on whether $Q$
transforms as a $\bf 2$ or a $\bf\bar 2$ under $U(2)$. The antiquark $u^c$
corresponds to fluctuations of an open string with one end on the color
branes and the other on the $U(1)$ brane for $x=\pm 1$, or on other branes in
the bulk for $x=0$. Ditto for $d^c$. Similarly, the lepton doublet $L$
arises from an open string with one end on the weak branes and the other
on the $U(1)$ brane for $z=\pm 1$, or in the bulk for $z=0$. Finally, $l^c$
corresponds necessarily to an open string with one end on the $U(1)$ brane and
the other in the bulk. We defined its $Q_1=1$.

The weak hypercharge $Y$ is a linear combination of the three
$U(1)$'s:\footnote{A study of hypercharge embeddings in gauge groups obtained
from M-branes was considered in Ref. \cite{west}.
In the context of Type I groundstates such embeddings were considered in 
\cite{ib}.}
\be
Y=c_1 Q_1+c_2 Q_2+c_3 Q_3\, .
\label{Y}
\ee
$c_1=1$ is fixed by the charges of $l^c$ in eq.~(\ref{charges}), while
for the remaining two coefficients and the unknown charges $x,y,z,w$, we obtain
four possibilities:
\bea
c_2 =-{1\over 2}\, ,\, c_3=-{1\over 3}\, ;&
x=-1\, ,\, y=0\, ,\, z=0\, ,\, w=-1\nonumber\\
c_2 ={1\over 2}\, ,\, c_3=-{1\over 3}\, ;&
x=-1\, ,\, y=0\, ,\, z=-1\, ,\, w=1\nonumber\\
c_2 =-{1\over 2}\, ,\, c_3={2\over 3}\,  ;&\!\!\!\!\!\!\!\!\!
x=0\, ,\, y=1\, ,\, z=0\, ,\, w=1\label{solutions}\\
c_2 ={1\over 2}\, ,\, c_3={2\over 3}\, ;&
x=0\, ,\, y=1\, ,\, z=-1\, ,\, w=-1\nonumber
\eea
Orientifold models realizing the $c_3=-1/3$ embedding in the supersymmetric case
with intermediate string scale $M_s\sim 10^{11}$ GeV have been described in 
\cite{ib}.

To compute the weak angle $\sin^2\theta_W$, we use from eq.~(\ref{Y}) that the
hypercharge coupling $g_Y$ is given by~\footnote{The gauge couplings
$g_{2,3}$ are determined at the tree-level by the string coupling and other
moduli, like radii of longitudinal dimensions. In higher orders, they also
receive string threshold corrections.}:
\be
{1\over g_Y^2}={2\over g_1^2}+{4c_2^2\over g_2^2}+
{6c_3^2\over g_3^2}\, ,
\label{gY}
\ee
with $g_1=g_2$ or $g_1=g_3$ at the string scale. On the other hand, with the
generator normalizations employed above, the weak $SU(2)$ gauge coupling is
$g_2$. Thus,
\be
\sin^2\theta_W\equiv{g_Y^2\over g_2^2+g_Y^2}=
{1\over 1+4c_2^2+2g_2^2/g_1^2+6c_3^2g_2^2/g_3^2}\, ,
\label{sintheta}
\ee
which for $g_1=g_2$ reduces to:
\be
\sin^2\theta_W(M_s)=
{1\over 4+6c_3^2g_2^2(M_s)/g_3^2(M_s)}\, ,
\label{sintheta12}
\ee
while for $g_1=g_3$ it becomes:
\be
\sin^2\theta_W(M_s)=
{1\over 2+2(1+3c_3^2)g_2^2(M_s)/g_3^2(M_s)}\, .
\label{sintheta13}
\ee

We now show that the above predictions agree with the experimental value for
$\sin^2\theta_W$ for a string scale in the region of a few TeV. For this
comparison, we use the evolution of gauge couplings from the weak scale $M_Z$ as
determined by the one-loop beta-functions of the Standard Model with three
families of quarks and leptons and one Higgs doublet,
\be
{1\over \alpha_i(M_s)}={1\over \alpha_i(M_Z)}-
{b_i\over 2\pi}\ln{M_s\over M_Z}\ ; \quad i=3,2,Y
\ee
where $\alpha_i=g_i^2/4\pi$ and $b_3=-7$, $b_2=-19/6$, $b_Y=41/6$. We also use
the measured values of the couplings at the $Z$ pole 
$\alpha_3(M_Z)=0.118\pm 0.003$, $\alpha_2(M_Z)=0.0338$, $\alpha_Y(M_Z)=0.01014$
(with the errors in $\alpha_{2,Y}$ less than 1\%). 

In order to compare the theoretical relations for the two cases
(\ref{sintheta12}) and (\ref{sintheta13}) with the experimental value of
$\sin^2\theta_W=g_Y^2/(g_2^2+g_Y^2)$ at $M_s$, we plot in Fig.~1
the corresponding curves as functions of $M_s$. 
%
\begin{figure}[htb]
\epsfxsize=6.5in
\epsfysize=5in
\hspace{1.5cm}
\epsffile{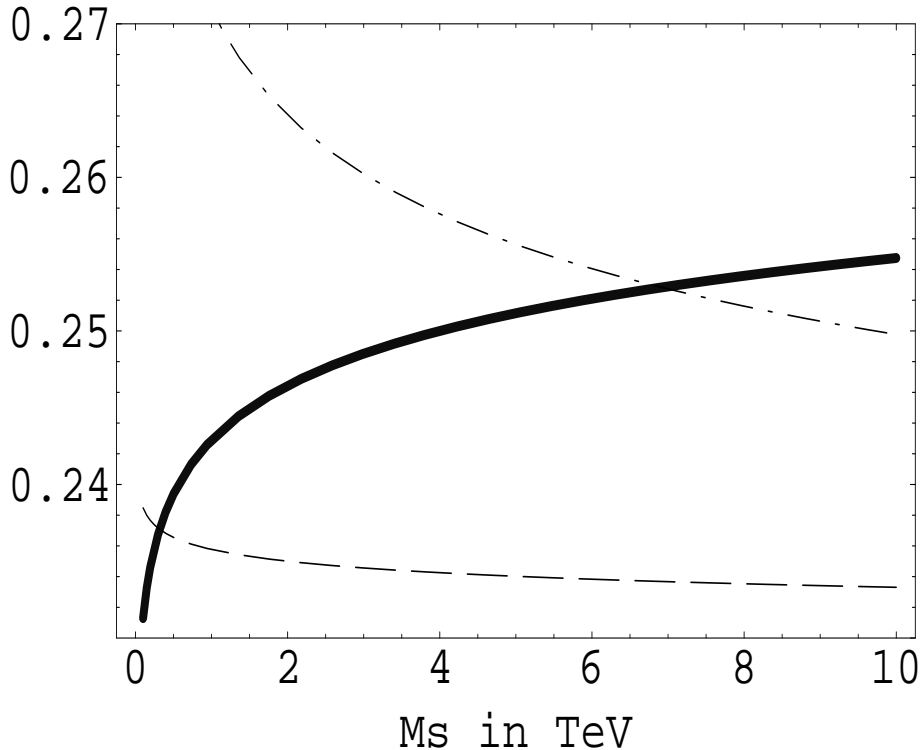}
\vspace{-2.5cm}
\caption{The experimental value of $\sin^2\theta_W$ (thick curve), together with
the theoretical predictions (\ref{sintheta12}) with $c_3=-1/3$ (dashed line) and
(\ref{sintheta13}) with $c_3=2/3$ (dotted-dashed), are plotted as functions of 
the string scale $M_s$.}
\label{sin}
\end{figure}
The solid line is the experimental curve. The dashed line is the plot of the
function (\ref{sintheta12}) for $c_3=-1/3$ while the dotted-dashed line
corresponds to the function (\ref{sintheta13}) for $c_3=2/3$. Thus, the second
case, where the $U(1)$ brane is on top of the color branes, is compatible with
low energy data for $M_s\sim 6-8$ TeV and $g_s\simeq 0.9$.
 This selects the last two possibilities of
charge assignments in Eq.~(\ref{solutions}). The curve corresponding to
$g_1=g_3$ and $c_3=-1/3$ gives for $M_s$  few thousand TeV. 
This value is too high to protect the hierarchy.
The other case, where the $U(1)$ brane is on top of the weak branes, is
not interesting either. For $c_3=2/3$, the corresponding curve does not intersect
the experimental one at all and is not shown in the Fig.~1, while the case of
$c_3=-1/3$ leads to $M_s$ of a few hundred GeV and is  excluded
experimentally. In the sequel we shall restrict ourselves to the last two
possibilities of Eq.~(\ref{solutions}).

{}From the general solution (\ref{solutions}) and the requirement that the Higgs
doublet has hypercharge $1/2$, one finds the following possible assignments for
it, in the notation of Eq.~(\ref{charges}):
\bea
\label{H}
c_2=-{1\over 2}\ :\qquad &H\ \ ({\bf 1},{\bf 2};0,1,1)_{1/2}\quad
&H'\ \ ({\bf 1},{\bf 2};0,-1,0)_{1/2}\\
c_2={1\over 2}\ :\qquad &{\tilde H}\ \ ({\bf 1},{\bf 2};0,-1,1)_{1/2}\quad
&{\tilde H}'\ \ ({\bf 1},{\bf 2};0,1,0)_{1/2}
\label{Htilde}
\eea
It is straightforward to check that the allowed (trilinear) Yukawa terms are:
\bea
\label{HY}
c_2=-{1\over 2}\ :\qquad H'Qu^c\ ,\quad H^\dagger Ll^c
\ ,\quad H^\dagger Qd^c\\
c_2={1\over 2}\ :\qquad {\tilde H}' Qu^c\ ,\quad {\tilde H}'^\dagger Ll^c
\ ,\quad {\tilde H}^\dagger Qd^c
\label{HtildeY}
\eea
Thus, two Higgs doublets are in each case necessary and sufficient to give masses
to all quarks and leptons. Let us point out that the presence of the second Higgs
doublet changes very little the curves of Fig.~1 and consequently our previous
conclusions about $M_s$ and $\sin^2\theta_W$.

A few important comments are now in order:

\noindent
(i) The spectrum we assumed in Eq.~(\ref{charges}) does not contain right-handed
neutrinos on the branes. They could in principle arise from open strings in the
bulk. Their interactions with the particles on the branes would then be
suppressed by the large volume of the transverse space \cite{Rnus}. 
More specifically, conservation of the three U(1) charges allow for the
following Yukawa couplings involving the right-handed neutrino $\nu_R$:
\be
c_2=-{1\over 2}\;:\;\;\; H'~L~{\nu_L}\;\;\;\;\;\;;\;\;\;\;\;\; c_2={1\over 2}\;:\;\;\;
\tilde H~ L ~\nu_R
\ee
These couplings lead to Dirac type neutrino masses between $\nu_L$ from $L$ and
the zero mode of $\nu_R$, which is naturally suppressed by the volume of the
bulk.
 
\noindent
(ii) Implicit in the above was our assumption of three generations (\ref{charges})
of quarks and lepton in the light spectrum. They can arise, for example, from
an orbifold action along the lines of the model described in Ref.~\cite{ib}.

\noindent
(iii) From Eq.~(\ref{sintheta13}) and Fig.~1, we find the ratio of the $SU(2)$ and
$SU(3)$ gauge couplings at the string scale to be $\alpha_2/\alpha_3\sim 0.4$. This
ratio can be arranged by an appropriate choice of the relevant moduli. For
instance, one may choose the color and U(1) branes to be D3 branes while 
the weak branes to be D7 branes.
Then the ratio of couplings above can be explained by choosing the volume of the four 
compact dimensions of the seven branes to be $V_{4}=2.5$ in string units.
This being larger than one is consistent with the picture above.
Moreover it predicts an interesting spectrum of KK states for the Standard model, different from 
the naive  that have appeared hitherto: The only Standard Model particles that
have KK descendants are the W bosons as well as the hypercharge gauge boson.
However since the hypercharge is a linear combination of the three U(1)'s the massive U(1) gauge 
bosons couple not to hypercharge but to doublet number.

Another
possibility would be to move slightly off the orientifold point which may be
necessary also for other reasons (see discussion towards the end of the paper).

\noindent
(iv) Finally, it should be stressed that 
there are some alternative assignments that may work and these are discussed further in 
\cite{akt}.

\section{The fate of $U(1)$'s and proton stability}

The model under discussion has three $U(1)$ gauge interactions corresponding to
the generators $Q_1$, $Q_2$, $Q_3$. From the previous analysis, the hypercharge
was shown to be either one of the two linear combinations:
\be
Y=Q_1\mp{1\over 2} Q_2+{2\over 3}Q_3\, .
\ee 
It is easy to see that the remaining two $U(1)$ combinations orthogonal to $Y$ are
anomalous.
In particular there are mixed anomalies with the SU(2) and SU(3) gauge groups of the Standard Model.
 
These anomalies are canceled by two axions coming from the closed string sector, via the standard Green-Schwarz
mechanism \cite{gs}.
The mixed anomalies with the non-anomalous hypercharge are also canceled by dimension five Chern-Simmons 
type of interactions \cite{akt}.
The presence of such interactions has so far escaped attention in the context of string theory.

An important property of the above Green-Schwarz anomaly cancellation mechanism
is that the two $U(1)$ gauge bosons $A$ and $A'$ acquire masses leaving behind
the corresponding global symmetries \cite{gs}. This is in contrast to
what would had happened in the case of an ordinary Higgs mechanism. These global
symmetries remain exact to all orders in type I string perturbation theory
around the orientifold vacuum.

So, as long as we stay at the orientifold point, all three charges $Q_1$, $Q_2$,
$Q_3$ are conserved and since $Q_3$ is the baryon number, proton stability is
guaranteed.

To break the electroweak symmetry, the Higgs doublets in Eq.~(\ref{H}) or
(\ref{Htilde}) should acquire non-zero VEV's. Since the model is
non-supersymmetric, this may be achieved radiatively \cite{abq}. From
Eqs.~(\ref{HY}) and (\ref{HtildeY}), to generate masses for all quarks and leptons,
it is necessary for both Higgses to get non-zero VEV's. The baryon number
conservation remains intact because both Higgses have vanishing $Q_3$. However, the
linear combination which does not contain $Q_3$, will be
broken spontaneously, as follows from their quantum numbers in Eqs.~(\ref{H}) and
(\ref{Htilde}). This leads to an unwanted massless Goldstone boson of the
Peccei-Quinn type. The way out is to break this global symmetry explicitly,
by moving away from the orientifold point along the direction of the associated modulus
so that baryon number remains conserved.
Instanton effects in that case will generate the appropriate symmetry breaking couplings 
in the potential.

\section{A fifth force?}

As is obvious from the previous discussion in order to explain the ratio of the strong to the weak coupling we must assume that the U(2) gauge group arises from a D7 brane with four compact directions, with radii $R_i$, $i=1,2,3,4$ and 
$R_1R_2R_3R_4 \sim 2.5$ and $R_i\geq 1$ in string units.
The other interactions arise from D3 branes.
By looking again on the charge assignments of open strings that represent
the fermions,  it is obvious that the $\bar U$ and the electron singlet 
(as well as one of the Higgses) must terminate in an extra set of branes different from the one described above.
In general, the gauge group of such a brane will be $U(n)$ for some $n\geq 1$.
The U(1) factor is necessarily anomalous. It is easy to check that it has mixed anomalies with the SU(3) of color.
Thus, this gauge boson will acquire a mass $M=g M_s$ where $g$ 
is the gauge coupling  of the anomalous U(1).
If such an extra set consists of D3 branes, or D7 branes parallel to the U(2)
branes, then this implies the existence of a new force among u-quarks and electrons with the same strength as the strong or the weak force.
The non-abelian piece SU(n), if there, must be broken completely 
at the string scale, otherwise it would be incompatible with data.
The U(1) is broken by the anomaly and the associated gauge boson has a mass of the order of TeV.

The alternative possibility is that the branes are D7 branes intersecting the U(2) branes along the 1,2
directions, the effective gauge coupling is $g^2=(4\pi \alpha_{\rm strong})^{-1}{M_s^2\over M_P^2}\sim 5\times 10^{-31}$, and the mass of the anomalous gauge boson is $M=g M_s\sim 5\times 10^{-3}~$eV which corresponds 
to a range of 40 $\mu$m. This is currently allowed by ``fifth force" data \cite{5} but is probably excluded by supernova physics due to the large cross section for producing such gauge bosons.\footnote{There is an alternative possibility, \cite{akt}, where the electron singlet is a string with both end on the weak brane. In this case the experimental constraints are weaker, since only the ${\bar u}$-quark feels the fifth force.}  
The gauge boson above,would be  the only particle with low lying KK excitations quantized in units of $~8\times 10^{-3}$ eV.
Low-lying KK states coming from the bulk have masses also quantized in 
units of $~8\times 10^{-3}$ eV and are thus of the same order as the masses described above.

In conclusion, we presented a particular embedding of the Standard Model in a
non-supersymmetric D-brane configuration of type I/I$^\prime$ string theory. The
strong and electroweak couplings are not unified because strong and weak
interactions live on different branes. Nevertheless, $\sin^2\theta_W$ is naturally
predicted to have the right value for a string scale of the order of a few TeV. The
model contains two Higgs doublets needed to give masses to all quarks and leptons,
and preserves baryon number as a (perturbatively) exact global symmetry. The model
satisfies the main phenomenological requirements for a viable low energy theory
and its explicit derivation from string theory deserves further study.

\vskip0.5cm
\noindent
{\large \bf Acknowledgements}\smallskip

\noindent
EK would like to thank the organizers of the RTN meeting "Quantum spacetime"
for the hospitality and excellent organization.
This work was partly supported by the EU under contracts ERBFMRX-CT96-0090, HPRN-CT-2000-00122, HPRN-CT-2000-00131 and INTAS contract  N 99 0590.


\end{document}